\documentclass{PoS}
\usepackage{bm}

\title{Charmed Meson Scattering from Lattice QCD}

\ShortTitle{Charmed Meson Scattering from Lattice QCD}

\author{\speaker{Graham Moir} \\
Department of Applied Mathematics and Theoretical Physics, Centre for Mathematical Sciences, University of Cambridge, Wilberforce Road, Cambridge, CB3 0WA, UK \\
E-mail: \email{graham.moir@damtp.cam.ac.uk}  \\

(For the Hadron Spectrum Collaboration)\\}

\bigskip

\abstract{State-of-the-art lattice QCD calculations of scattering amplitudes in coupled-channel $D\pi$, $D\eta$ and $D_{s}\bar{K}$ scattering, as well elastic $DK$ scattering are discussed. The methodology employed allows a determination of the relevant poles in the scattering matrix, while also providing a measure of the coupling of each channel to a given pole. By investigating $S$, $P$ and $D$ wave interactions, the nature of states with $J^{P} = 0^{+}$, relevant for the $D^{*}_{0}(2400)$ and $D^{*}_{s0}(2317)$, as well as states with $J^{P} = 1^{-}, 2^{+}$ are discussed.}

\FullConference{VIII International Workshop On Charm Physics\\
5-9 September, 2016\\
Bologna, Italy \hspace{10.5cm} DAMTP-2016-75}

\begin{document}

\section{Introduction}
\label{intro}

Since the turn of the century, there has been a rapid increase in the number of observed resonances appearing to have valence charm degrees of freedom \cite{PDG}. The most theoretically interesting are those that do not adhere to expectations coming from conventional quark models. Since the majority of these appear as charmonium-like resonances high up in the spectrum above many multi-meson thresholds, they can be difficult to study in lattice calculations. One notable exception however, due to its appearance low down in the spectrum of mesons containing charm quarks, is the enigmatic $D^{*}_{s0}(2317)$ whose qualitative differences with the $D^{*}_{0}(2400)$ are not yet understood. In quark potential models, they are both expected to behave in a similar way, that is, to appear as broad resonances above the $D\pi$ and $DK$ thresholds respectively. Although the $D^{*}_{0}(2400)$ conforms to this expectation, the $D^{*}_{s0}(2317)$ does not; experimentally it is found below the $DK$ threshold and is extremely narrow. 

In these proceedings, I provide a summary of the coupled-channel $D\pi$, $D\eta$ and $D_{s}\bar{K}$ scattering calculation presented in Ref.~\cite{MOIR} which is relevant for the $D^{*}_{0}(2400)$, along with an initial preliminary study of elastic $DK$ scattering which is relevant for the $D^{*}_{s0}(2317)$. Previous lattice calculations of elastic $D\pi$ and $DK$ scattering performed by other groups are discussed in Refs.~\cite{LANG, MOHLER, MOHLER2}. 

The calculations described here are performed on dynamical $N_{f} = 2 + 1$ ensembles generated by the Hadron Spectrum Collaboration in which the spatial lattice spacing $a_{s}$ is roughly $3.5$ times larger than the temporal lattice spacing $a_{t}$. The same relativistic action is employed for the light, strange and (valence) charm quarks where the charm quark has been tuned using the physical $\eta_{c}$ mass \cite{LIU}. Note that $m_{\pi} = 391$ MeV on these ensembles. Further details of the lattice set-up are given in Ref.~\cite{EDWARDS, LIN}.

\section{Scattering on the lattice}
\label{scatt}

In Minkowski space-time, resonances appear as poles on unphysical Riemann sheets of the scattering $S$-matrix. However, lattice calculations are currently performed in a finite-volume of Euclidean space-time, which has major implications for the study of scattering and resonances as the finite volume renders all particles stable. This is a consequence of the loss of the multi-particle branch cut which dissolves into a series of poles on the physical sheet; since there is no branch cut, there is no access to the unphysical sheets. One might consider attempting to construct some appropriate $n$-point function which directly contains infinite-volume scattering information, however, the Maiani-Testa theorem forbids the existence of such quantities in Euclidean space-time as they lack the required non-trivial complex phase~\cite{MAIANI}.

\subsection{Obtaining scattering amplitudes}

Obtaining scattering information `indirectly' via the use of the L\"{u}scher formalism~\cite{LUSCHER, LUSCHER2} has become the norm, with more recent calculations making use of the the extension of the method to moving frames~\cite{RUMMUKAINEN,KIM,LESKOVEC,FU} and to coupled-channels~\cite{HANSEN2,BRICENO2,GUO,LIU2}. In this approach the finite volume of the lattice is used as a tool and exponentially suppressed corrections, which fall off like $e^{-m_{\pi}L}$, where $L$ is the spatial extent of the lattice, are neglected. These unwanted corrections are known to be negligible for values of $m_{\pi}L > 4$; the volumes used in this work have $m_{\pi}L \sim 4$ to $6$.

The key feature of this formalism is that it relates the spectrum determined in a finite volume to the infinite-volume scattering matrix. Therefore, at least in principle, one should be able to input the spectroscopic energies determined from a lattice QCD calculation, performed in a finite volume of Euclidean space-time, and extract infinite-volume scattering amplitudes. For two-body scattering, the main difficulty arises from the fact that for $N$ coupled-channels, the non-trivial part of the scattering matrix (referred to here as the $t$-matrix) has $(N^{2} + N)/2$ unknowns and hence, the problem is under-constrained for all but the single-channel case ($N = 1$). One way to proceed is to parametrise the energy dependence of the $t$-matrix using a relatively small number of free parameters. This parametrisation can then be constrained globally using a large number of finite-volume energies determined in a lattice QCD calculation. Once a parametrised $t$-matrix is in hand, it can be analytically continued into the complex energy plane where its pole content can be explored. 

In order to reduce the dependence on the choice of parametrisations used, this process is repeated for a variety of parametrisations; those that can successfully reproduce the finite-volume lattice QCD spectra  are incorporated into the final results. The parametrisations employed here in the case of elastic scattering are the effective range expansion, the relativistic Breit-Wigner and various forms of the $K$-matrix. For coupled-channel scattering, various forms of the $K$-matrix are employed. All parametrisations used are described in detail in Refs.~\cite{MOIR, WILSON}.

\subsection{Obtaining finite-volume spectra}

As indicated above, the extraction of scattering amplitudes relies on the ability to determine relevant spectroscopic energies in a finite-volume. This is achieved through the computation of Euclidean two-point  functions 
\begin{equation}
C_{ij}(t) = \left< 0 \left| \mathcal{O}_{i}(t) \mathcal{O}^{\dagger}_{j}(0) \right| 0 \right>~,
\end{equation}
whose time-dependence is related to energies of the states created by the interpolating operator $\mathcal{O}^{\dagger}$. Using a large basis of well-chosen interpolating operators, a robust determination of many finite-volume energies can be obtained by solving a generalised eigenvalue problem for  $C_{ij}(t)$. 

\begin{figure}[t!]
\centering
\includegraphics[width=0.75\textwidth]{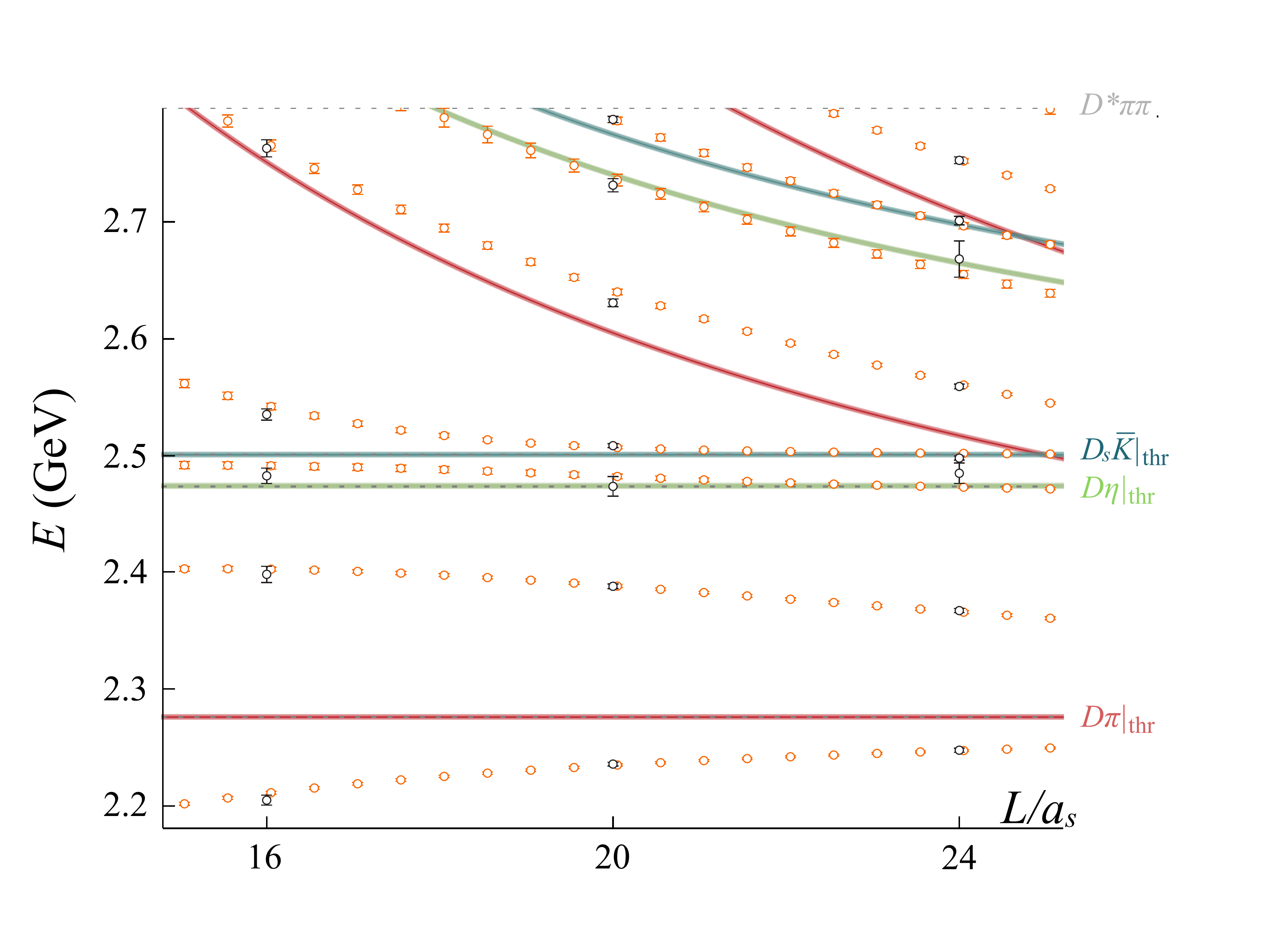}
\caption{The spectrum (black points) in a lattice symmetry channel ($A^{+}_{1}$) dominated by $J^{P}=0^{+}$ interactions for three lattices of different spatial extents $L/a_{s}$. The solid curves correspond to non-interacting $D\pi$ (red), $D\eta$ (green) and $D_{s}\bar{K}$ (blue) energies for various ``back-to-back'' momenta resulting in zero overall momentum of the system. The orange points correspond to the spectrum coming from an example $K$-matrix parametrisation deemed able to reproduce the lattice QCD spectrum.}
\label{fig1} 
\end{figure}

As an example, Fig.~\ref{fig1} shows the finite-volume energies (black points) determined on three lattices with different spatial volumes $(L/a_{s})^{3}$ using various $D\pi$-like,  $D\eta$-like,  $D_{s}\bar{K}$-like and $D$ meson-like operators. The solid lines represent the non-interacting energies of $D\pi$ (red), $D\eta$ (green) and $D_{s}\bar{K}$ (blue) combinations for various ``back-to-back'' momenta resulting in zero overall momentum of the system. Note that, the lattice symmetry channel shown ($\Lambda^{P} = A^{+}_{1}$) is dominated by $J^{P}=0^{+}$ contributions.

\section{Coupled-channel $D\pi$, $D\eta$ and $D_{s}\bar{K}$ scattering}

Once the lattice QCD spectrum is obtained in all relevant symmetry channels, it is used to constrain a variety of parametrisations of the $t$-matrix as a function of energy. This is achieved using a $\chi^{2}$ function that minimises the difference between the lattice QCD spectrum and the parametrised spectrum obtained via the L\"{u}scher formalism. The result of one such minimisation is shown in Fig.~\ref{fig1}; the black points correspond to the lattice QCD spectrum and the orange points are those coming from an example $K$-matrix parametrisation that clearly reproduces the lattice QCD spectrum. Note that, the minimisation is not only to the energies shown in this lattice symmetry channel but to all relevant symmetry channels including those with overall non-zero momentum of the system. This is crucial in order to constrain the $t$-matrix across the entire energy range of interest and help prevent the appearance of spurious poles in the parametrised $t$-matrices. This process is repeated for a large variety of parametrisations of the $K$-matrix (see Table 11 of Ref. \cite{MOIR}) and all those deemed capable of reproducing the lattice QCD spectrum are included in the uncertainty quoted for the final results.

\begin{figure}[t!]
\centering
\includegraphics[width=0.7\textwidth]{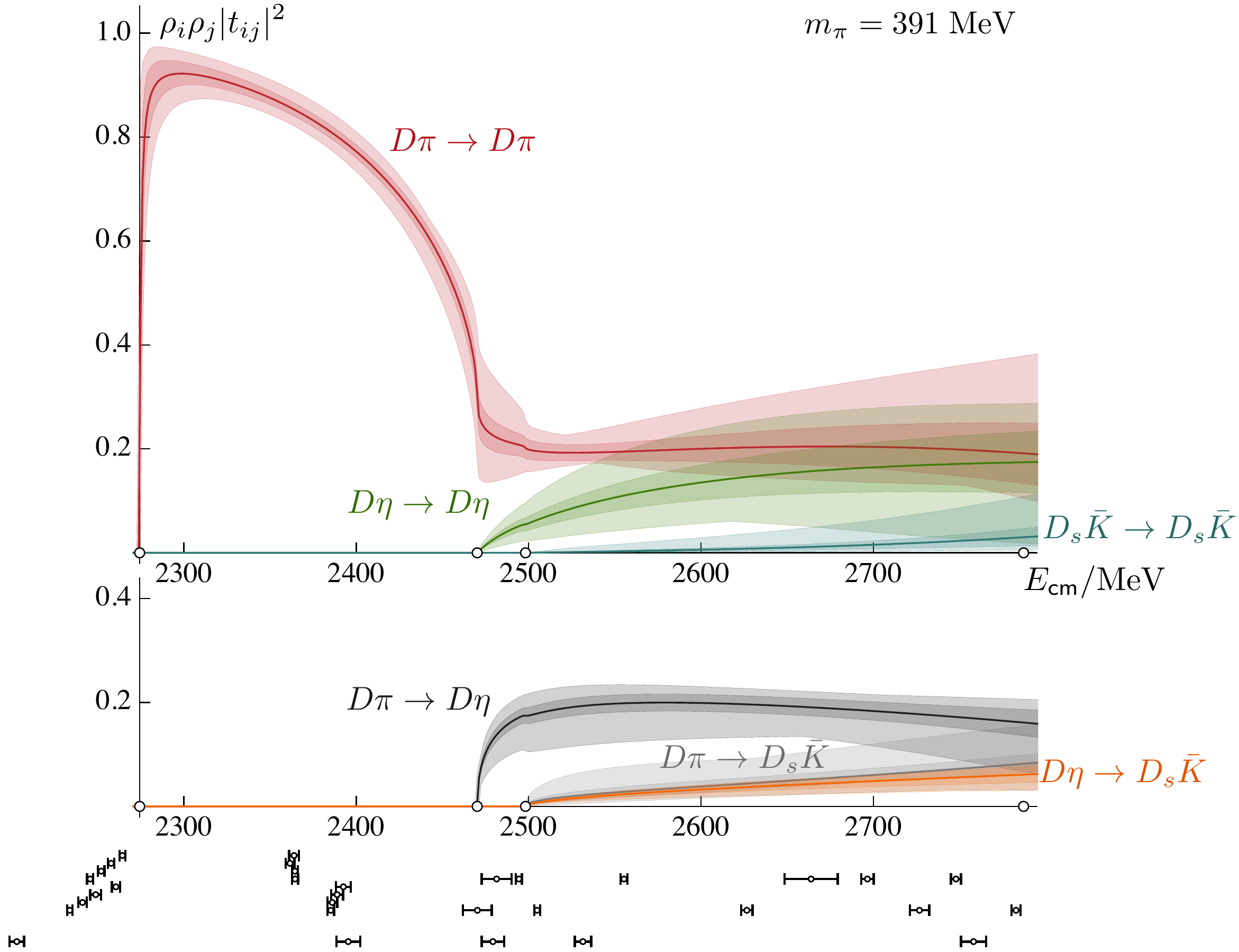}
\caption{The quantity $\rho_{i}\rho_{j}|t_{ij}|^{2}$ for the $S$-wave part of the $t$-matrix. The upper panel shows the diagonal elements while the lower panel shows the off-diagonal elements. The colour coding is as indicated in the figure and the width of each band represents the statistical uncertainty as well as the systematic uncertainty coming from the $K$-matrix parametrisations that were able to reproduce the lattice QCD spectrum. From left to right, the open circles on the horizontal axes correspond to location of the $D\pi$, $D\eta$, $D_{s}\bar{K}$ and $D^{*}\pi\pi$ thresholds respectively and the black points below the lower panel show the location of the lattice QCD energies used to constrain the $t$-matrix.}
\label{fig2} 
\end{figure}

An enlightening way of presenting the results is to look at a quantity proportional to the cross-section,  $\rho_{i}\rho_{j}|t_{ij}|^{2}$, where the phase-space factor $\rho_{i} = 2k_{i}/E_{cm}$ and $t_{ij}$ are elements of the scattering $t$-matrix. This is shown for the $S$-wave part of the $t$-matrix in Fig.~\ref{fig2}. The $D\pi \rightarrow D\pi$ amplitude shows a large "peak" almost saturating the unitarity bound (which  is unity) just above the $D\pi$ threshold - a clear indication of non-trivial interaction.

To take the analysis one step further, an interpretation of the scattering amplitudes is required. To this end, the parametrised $t$-matrices are now considered functions of complex energies where their pole contents can be associated with (virtual) bound-states and resonances. In all of the parametrised $t$-matrices able to reproduce the lattice QCD spectra, a bound-state pole in the $S$-wave part of the $t$-matrix is consistently found just below the $D\pi$ threshold. Averaging over the location of the poles from these parametrisations leads to a final result of $2275.9 \pm 0.9$ MeV, where the uncertainty encompasses the uncertainty from the individual parametrisations. This pole, which is statistically indistinguishable from the $D\pi$ threshold appears to be solely responsible for the broad peak appearing in the $D\pi \rightarrow D\pi$ amplitude in Fig.~\ref{fig2}. Within the considered energy region, no other $S$-wave pole appeared consistently across all parametrisations that were able to reproduce the lattice QCD spectra. 

\begin{figure}[t!]
\centering
\includegraphics[width=0.7\textwidth]{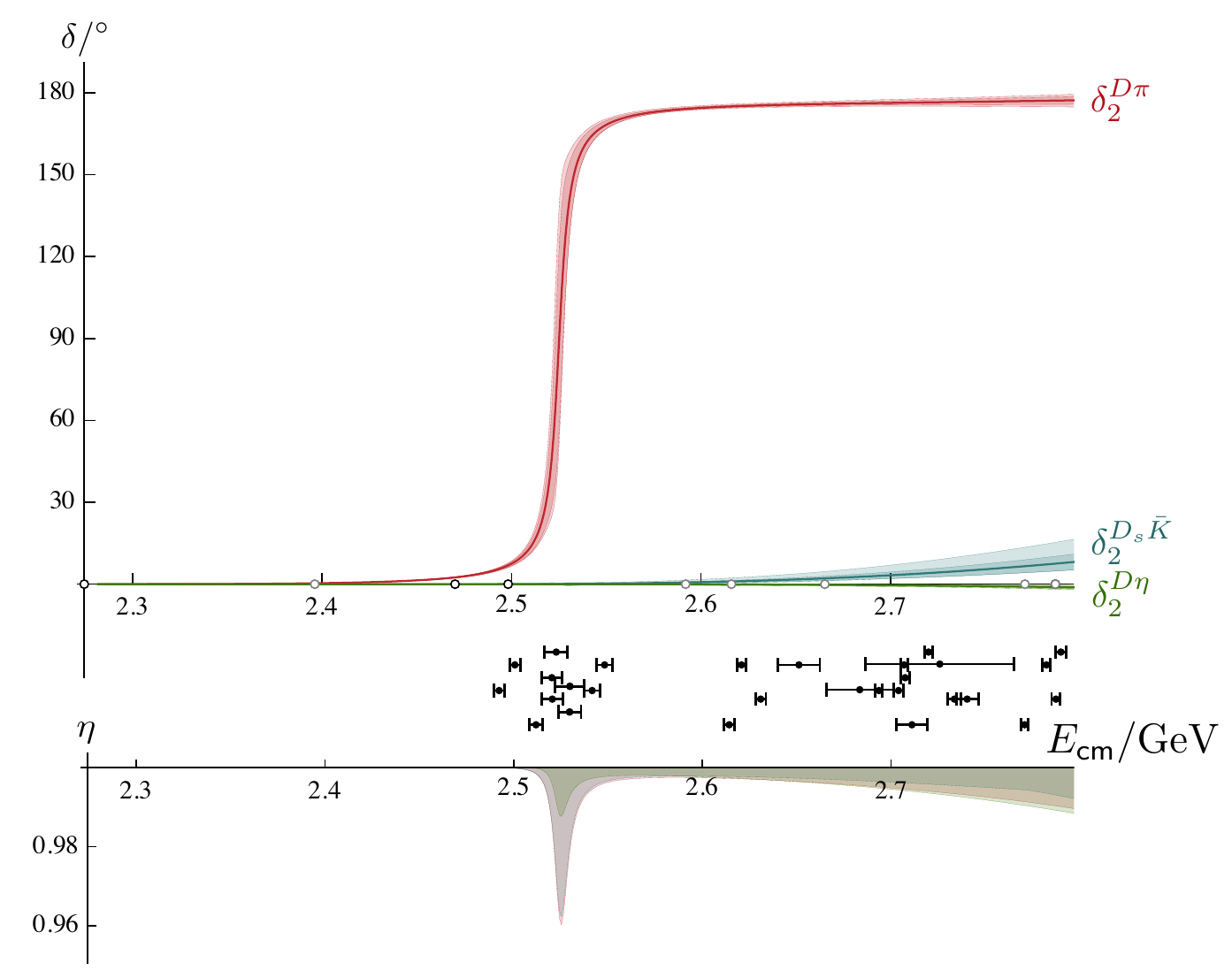}
\caption{The upper panel shows the $D$-wave  $D\pi$ (red), $D\eta$ (green) and $D_{s}\bar{K}$ (blue) phase-shifts. The lower panel shows the inelasticities $\eta_{i}$ taken from the diagonal parts of the $t$-matrix. The black points between the panels show the location of the finite-volume energies used to constrain the scattering amplitudes.}
\label{fig3} 
\end{figure}

These parametrisations also simultaneously considered symmetry channels that included both $S$ and $P$-wave contributions, and hence, kept the unphysical mixing  between the $\ell = 0$ and the $\ell = 1$ partial waves introduced by the finite volume under control. In all parametrisations able to reproduce the lattice QCD spectra, a $P$-wave bound-state pole was observed far below the $D\pi$ threshold. Averaging over these parametrisations results in a final result of $2009 \pm 2$ for the location of the pole where again, the uncertainty encompasses the uncertainty from the individual parametrisations.  This pole lies almost exactly at the same energy as the experimental $D^{*}(2007)$; a very narrow state whose mass is $2006.85 \pm 0.05$ MeV~\cite{PDG}. Since this pole is located far below the $D\pi$ threshold it is not expected to strongly influence $D\pi$ scattering, which is consistent with the small amplitudes obtained in $P$-wave.

Parametrisations for the $D$-wave part of the $t$-matrix were obtained using finite-volume energies whose dominant contribution comes from $\ell=2$. 	The upper panel of Fig.~\ref{fig3} shows the resulting  $D\pi$ (red), $D\eta$ (green) and $D_{s}\bar{K}$ (blue) phase-shifts. The rapid shift through $180^{\circ}$ in the $D\pi$ channel is a classic sign of a resonant interaction. Examining the pole structure yields a resonance pole on all sheets with $\mathrm{Im}[k_{D{\pi}}] < 0$ in all parametrisations able to reproduce the lattice QCD spectrum. Averaging over the location of the pole on all sheets and parametrisations gives a pole mass and width of $2527 \pm 3$ MeV and $8.2 \pm 0.7$ MeV respectively. Although this $J^{P}=2^{+}$ resonance appears somewhat different to the experimental $D^{*}_{2}(2460)$, whose mass and width are $2460.57 \pm 0.15$ MeV and $47.7 \pm 1.3$ MeV~\cite{PDG} respectively, it is fully expected to approach the experimental values as the pion mass is decreased.

\section{$DK$ scattering and the $D_{s0}(2317)$}

Following the procedure described in section 2, finite-volume energies were obtained by solving a generalised eigenvalue problem for a matrix of two-point correlation functions built from a variational basis that includes $DK$-like and $D_{s}$-like interpolating operators. The scattering analysis that followed only used finite-volume energies lying below the $D_{s}\eta$ threshold resulting in parametrised $t$-matrices for elastic $DK$ scattering. 

On examining the $S$-wave part of these matrices, a bound-state pole approximately $55$ MeV below the $DK$ threshold was consistently found in the parametrisations able to reproduce the lattice spectrum.  Although this preliminary calculation is performed at an unphysical pion mass of $m_{\pi} = 391$ MeV, it is interesting to compare with the experimental $D^{*}_{s0}(2317)$ whose mass is roughly $45$ MeV below the physical $DK$ threshold~\cite{PDG}.

\section{Summary}

Scattering amplitudes obtained from coupled-channel $D\pi$, $D\eta$ and $D_{s}\bar{K}$ scattering along with those from a preliminary study of elastic $DK$ scattering were discussed. Analytic continuation of these amplitudes into the complex plane has allowed for a determination of their pole content. At $m_{\pi} = 391$ MeV, three poles were found in coupled-channel $D\pi$, $D\eta$ and $D_{s}\bar{K}$ scattering. In $S$-wave, a $J^{P} = 0^{+}$ bound-state pole, relevant for the $D^{*}_{0}(2400)$, was found just below the $D\pi$ threshold. In $P$-wave, a $J^{P} = 1^{-}$ bound-state pole was found to lie within a few MeV of the location of the $D^{*}(2007)$, while in $D$-wave, a $J^{P} = 2^{+}$ resonance pole relevant for the $D^{*}_{2}(2460)$ was found.

A single $S$-wave bound-state pole was observed in elastic $DK$ scattering. Its distance to the $DK$ threshold of roughly $55$ MeV is quite similar to the experimental case, where the mass of the $D^{*}_{s0}(2317)$ is approximately $45$ MeV below the experimental $DK$ threshold.

\section*{Acknowledgements}
GM thanks his colleagues within the Hadron Spectrum Collaboration and acknowledges support from the Herchel Smith Fund at the University of Cambridge.

\end{document}